\documentclass[a4paper, 12pt, openany, oneside]{article}
\usepackage[cp1251]{inputenc}
\usepackage[english, russian]{babel}
\usepackage{amsmath, latexsym,  amssymb, bm,  graphics, eucal,
amsfonts, array}
\usepackage[dvips]{graphicx}
\makeatletter

\newcommand{\rot}{\mathop{\rm rot\,}}
\makeatother
\usepackage{indentfirst}
\makeatother

\begin{document}
\renewcommand{\refname}{\begin{center} \bf References\end{center}}
\thispagestyle{empty}
\renewcommand{\abstractname}{Abstract}
\renewcommand{\contentsname}{Contents}
 \large
\newcommand{\mc}[1]{\mathcal{#1}}
\newcommand{\E}{\mc{E}}

 \begin{center}
\bf Nonlinear longitudinal current in quantum plasma
generated by two transversal electromagnetic waves
\end{center}\bigskip

\begin{center}
  \bf A. V. Latyshev\footnote{$avlatyshev@mail.ru$} and
  A. A. Yushkanov\footnote{$yushkanov@inbox.ru$}
\end{center}\medskip
\bigskip

\begin{center}
{\it Faculty of Physics and Mathematics,\\ Moscow State Regional
University, 105005,\\ Moscow, Radio str., 10-A}
\end{center}\medskip

\begin{abstract}
Quantum plasma with arbitrary degree degeneration of electronic gas
is considered. Plasma is in two external electromagnetic fields.
It is required to find the plasma response on this fields.
From Wigner kinetic equation  for the quantum
collisionless plasmas distribution function
in square-law approach on sizes of vector potentials
of two electric fields is received.
The formula for calculation electric current at any temperature
(any degree of degeneration of electronic gas) is deduced.
This formula contains an one-dimensional quadrature.
It is shown, that the nonlinearity account leads to occurrence
the longitudinal electric current directed along a wave vectors.
This longitudinal current is orthogonal to a known transversal current,
received at the linear analysis.
The case of small values of wave numbers is considered. It is shown, that
in case of small values of wave numbers a longitudinal current in
quantum plasma coincides with a longitudinal current in  classical
plasma.

{\bf Key words:} quantum plasma, Wigner equation, vector
potential, electromagnetic fields, current density,
transversal current, longitudinal current.

PACS numbers: 03.65.-w Quantum mechanics, 05.20.Dd Kinetic theory,
52.25.Dg Plasma kinetic equations.
\end{abstract}

\tableofcontents
\setcounter{secnumdepth}{4}

\begin{center}
\section*{Introduction}
\end{center}

В настоящей работе выводятся формулы для вычисления электрического
тока в квантовой бесстолкновительной плазме Ферми---Дирака.

В квантовой плазме выбираются четыре малых параметра.
Кинетическое уравнение Вигнера, описывающее поведение квантовой
плазмы, решается методом последовательных приближений.

При решении кинетического уравнения Вигнера мы учитываем как в разложении
функции распределения, так и в разложении  интеграла
Вигнера величины, пропорциональные
квадратам векторных потенциалов внешних электрических полей
и их произведению.

При таком нелинейном подходе оказывается, что электрический ток
имеет две ненулевые компоненты. Одна компонента электрического
тока направлена вдоль векторных потенциалов
электромагнитных полей.
Эта компонента  электрического тока точно такая же,
как и в линейном анализе. Это "поперечный"\,
ток.

Таким образом, в линейном приближении мы получаем известное
выражение поперечного электрического тока.

Вторая ненулевая компонента электрического тока имеет второй
порядок малости относительно малых параметров, пропорциональных
величинам напряженностей электрических полей.
Вторая компонента электрического тока направлена вдоль волнового
вектора. Этот ток ортогонален первой компоненте. Это "продольный"\, ток.

Появление продольного тока выявляется проведенным
нелинейным анализом взаимодействия электромагнитных полей
с плазмой.

Нелинейные эффекты в  плазме изучаются уже длительное время
\cite{Gins}--\cite{Lat1}.

В работах \cite{Gins} и \cite{Zyt} изучаются нелинейные эффекты в
плазме. В работе \cite{Zyt} нелинейный ток использовался, в
частности, в вопросах вероятности распадных процессов. Отметим,
что в работе \cite{Zyt2} указывается на существование
нелинейного тока вдоль волнового вектора (см. формулу (2.9) из \cite{Zyt2}).
В экспериментальной работе \cite{Akhmediev} выяснен вклад
нормальной компоненты поля в нелинейный поверхностный ток в
сигнале второй гармоники. В работах \cite{Urupin1, Urupin2}
изучалась генерация нелинейного поверхностного тока при
взаимодействии лазерного импульса с металлом.

Квантовая плазма изучалась в работах \cite{Lat1}--\cite{Lat9}.
Столкновительная квантовая плазма начала изучаться в работе
Мермина \cite{Mermin}. Затем квантовая столкновительная плазма
изучалась в наших работах \cite{Lat2}--\cite{Lat5}. Нами
изучалась квантовая столкновительная плазма с переменной частотой
столкновений. В работах    \cite{Lat7} -- \cite{Lat9} было
исследовано генерирование продольного тока поперечным
электромагнитным полем в классической и квантовой плазме
Ферми---Дирака \cite{Lat7} и в вырожденной плазме \cite{Lat9}.

Укажем еще на ряд работ по  плазме, в том числе и  квантовой.
Это работы \cite{Andres}--\cite{Manf}.

В настоящей работе выводятся формулы для вычисления электрического
тока в квантовой бесстолкновительной плазме
при произвольной температуре, то-есть при произвольной
степени вырождения электронного газа.

\section{Wigner equation}

Покажем, что в случае квантовой плазмы, описываемой уравнением
Вигнера, генерируется продольный ток и вычислим его плотность. На
существование этого тока указывалось более полувека тому назад
\cite{Zyt2}.

Будем считать, что квантовая плазма находится в двух внешних
электромагнитных полях с векторными потенциалами,
представляющими собой бегущие гармонические волны:
$$
{\bf A}_j({\bf r},t)={\bf A}_{0j}e^{i({\bf k}_j{\bf r}-\omega_jt)}
\quad (j=1,2).
$$

Соответствующие электрические и магнитные поля
$$
{\bf E}_j={\bf E}_{0j}e^{i({\bf k_jr}-\omega_j t)}, \qquad
{\bf H}_j={\bf H}_{0j}e^{i({\bf k_jr}-\omega_j t)}, \quad (j=1,2)
$$
связаны с векторными потенциалами равенствами
$$
{\bf E}_j=-\dfrac{1}{c}\dfrac{\partial {\bf A}_j}{\partial t}=
\dfrac{i\omega_j}{c}{\bf A}_j,\qquad
{\bf H}_j=\rot {\bf A}_j \qquad (j=1,2).
$$

Будем считать, что векторный потенциал электромагнитного поля
${\bf A}_j({\bf r},t)$ ортогонален волновому вектору ${\bf
k}_j$, т.е.
$$
{\bf k}_j\cdot {\bf A}_j({\bf r},t)=0,\qquad j=1,2.
$$
Это значит, что волновой вектор ${\bf k}_j$ ортогонален
электрическому и магнитному полям:
$$
{\bf k}_j\cdot {\bf E}_j({\bf r},t)=
{\bf k}_j\cdot {\bf H}_j({\bf r},t)=0,\qquad j=1,2.
$$

Для определенности будем считать, что волновые векторы обеих
полей направлены вдоль оси $x$, а электромагнитные поля
направлены вдоль оси $y$,
т.е.
$$
{\bf A}_j=A_j(x,t)(0,1,0),\qquad
A_j(x,t)=A_{0j}e^{i(k_jx-\omega_jt)},
$$
$$
{\bf k}_j=k_j(1,0,0), \quad {\bf E}_j=E_{j}(x,t)(0,1,0),\quad
E_j(x,t)=E_{0j}e^{i(k_jx-\omega_j t)}.
$$

Возьмем уравнение Вигнера, описывающее поведение квантовой
бесстолкновительной плазмы
$$
\dfrac{\partial f}{\partial t}+\mathbf{v}\dfrac{\partial f}{\partial
\mathbf{r}}+W[f]=0
\eqno{(1.1)}
$$
с нелинейным интегралом Вигнера
$$
W[f]= {\bf p}\sum\limits_{j=1}^{2}\dfrac{ie{\bf A}_j}{mc\hbar}
\Big[f\Big({\bf r},{\bf p}+\dfrac{\hbar{\bf k}_j}{2},t\Big)-
f\Big({\bf r},{\bf p}-\dfrac{\hbar{\bf k}_j}{2},t\Big)\Big]-
$$
$$
-\dfrac{ie^2}{2mc^2\hbar}\sum\limits_{j=1}^{2}{\bf A}_j^2
\Big[f\Big({\bf r},{\bf p+{\hbar k}}_j,t\Big)-
f\Big({\bf r},{\bf p-{\hbar k}}_j,t\Big)\Big]-
$$
$$
-\dfrac{2ie^2{\bf A}_1{\bf A}_2}{2mc^2\hbar}
\Big[f\Big({\bf r},{\bf p}+\hbar\dfrac{{\bf k}_1+{\bf k}_2}{2},t\Big)-
f\Big({\bf r},{\bf p}-\hbar\dfrac{{\bf k}_1+{\bf k}_2}{2},t\Big)\Big].
$$\medskip

Этот интеграл Вигнера построен по аналогии с интегралом Вигнера,
выведенным в работе \cite{Lat1} с использованием одного
векторного потенциала.

В уравнении (1.1) $f$ -- аналог квантовой функции распределения
электронов плазмы  (так называемая функция Вигнера),
$c$ -- скорость света,
${\bf p}=m{\bf v}$ -- импульс электронов,
${\bf v}$ -- скорость электронов.

Ниже нам понадобится локально равновесное распределение Ферми---Дирака,
$f^{(0)}=f_{eq}({\bf r},v)$ (eq$\equiv$ equi\-lib\-rium)

$$
f_{eq}({\bf r},v)=\Big[1+\exp\dfrac{\E-\mu({\bf
r})}{k_BT}\Big]^{-1},
$$
$\E={mv^2}/{2}$ -- энергия электронов, $\mu$ -- химический
потенциал электронного газа, $k_B$ -- постоянная Больцмана, $T$
-- температура плазмы,
$v_T$ -- тепловая скорость электронов,
$$
v_T=\sqrt{\dfrac{2k_BT}{m}},\qquad k_BT=\E_T=\dfrac{mv_T^2}{2},
$$
$\E_T$ -- тепловая кинетическая энергия электронов.

В квантовой плазме скорость электронов связана с его импульсом и
векторными потенциалами электромагнитных полей
равенством
$$
{\bf v}=\dfrac{{\bf p}}{m}-\dfrac{e}{cm}({\bf A}_1+{\bf A}_2).
$$

Если ввести безразмерный импульс электронов
${\bf P}={{\bf p}}/{p_T}$, где  $p_T=mv_T$ -- тепловой импульс
электронов, то предыдущее равенство перепишется в виде
$$
{\bf v}=v_T\Big({\bf P}-\dfrac{e}{cp_T}({\bf A}_1+{\bf
A}_2)\Big).
$$

Кроме того нам понадобится абсолютное распределение Ферми---Дирака
$f_0(p)$,
$$
f_0(p)=\Big[1+\exp\dfrac{p^2/2m-\mu}{k_BT}\Big]^{-1}=
\Big[1+\exp\Big(\dfrac{p^2}{p_T^2}-\alpha\Big)\Big]^{-1}=$$$$=
\dfrac{1}{1+e^{P^2-\alpha}}=f_0(P).
$$

Здесь
$\alpha=\mu/(k_BT)$ -- безразмерный химический потенциал
электронного газа.

В уравнении (1.1) и в интеграле Вигнера перейдем к безразмерному импульсу.
Получим уравнение
$$
\dfrac{\partial f}{\partial t}+v_T{\bf P}
\dfrac{\partial f}{\partial {\bf r}}+
 \dfrac{ie v_T{\bf P}}{c\hbar}\sum\limits_{j=1}^{2}{\bf A}_j
\Big[f\big({\bf r},{\bf P}+\dfrac{{\bf q}_j}{2},t\big)-
f\big({\bf r},{\bf P}-\dfrac{{\bf q}_j}{2},t\Big)\big]-
$$
$$
-\dfrac{ie^2}{2mc^2\hbar}\sum\limits_{j=1}^{2}{\bf A}_j^2
\Big[f({\bf r},{\bf p+{q}}_j,t)-
f({\bf r},{\bf p}-{\bf q}_j,t)\Big]-
$$
$$
-\dfrac{2ie^2{\bf A}_1{\bf A}_2}{2mc^2\hbar}
\Big[f\big({\bf r},{\bf p}+\dfrac{{\bf q}_1+{\bf q}_2}{2},t\big)-
f\big({\bf r},{\bf p}-\dfrac{{\bf q}_1+{\bf q}_2}{2},t\big)\Big].
\eqno{(1.2)}
$$

Здесь ${\bf q}_j$ -- безразмерные волновые числа,
${\bf q}_j={\bf k}_j/{k_T}$, $k_T$ -- тепловое волновое
число, $k_T=mv_T/\hbar$.

Будем искать решение уравнения (1.2) в виде
$$
f=f_0(P)+f_1+f_2,
\eqno{(1.3)}
$$
где
$$
f_1=A_1\varphi_1+A_2\varphi_2,
\eqno{(1.4)}
$$
$$
f_2=A_1^2\psi_1+A_2^2\psi_2+A_1A_2\psi_{0}.
\eqno{(1.5)}
$$

\section{First approximatuon of Wigner equation solution}

В задаче имеется четыре параметра размерности длины
$\lambda_j={v_T}/{\omega_j}$ ($v_T$ -- тепловая скорость электронов)
и $l_j={1}/{k_j}$.
Будем полагать, что как на длинах $\lambda_j$, так и на длинах $l_j$
изменение энергии электрона под действием соответствующего
потенциала электромагнитного поля $A_j$ много меньше
тепловой энергии электронов
$k_BT$ ($k_B$ -- постоянная Больцмана, $T$ -- температура плазмы),
т.е. будем считать малыми параметры
$$
\alpha_j=\dfrac{{\left|eA_j\right|v_T}}{c k_BT}\qquad (j=1,2)
$$
и
$$
\beta_j=\dfrac{{\left|eA_j\right|\omega_j}}{k_j k_BTc}
\qquad (j=1,2).
$$

Если воспользоваться связью векторных потенциалов
электромагнитных полей с напряженностями соответствующих
электрических полей, то введенные малые параметры выражаются
следующими равенствами
$$
\alpha_j=\dfrac{{\left|eE_j\right|v_T}}{\omega_j k_BT}\qquad (j=1,2)
$$
и
$$
\beta_j=\dfrac{{\left|eE_j\right|}}{k_j k_BT}\qquad (j=1,2).
$$

Будем действовать методом последовательных приближений, считая, что
$$
\alpha_j\ll 1 \qquad (j=1,2)
$$
и
$$
\beta_j\ll 1  \qquad (j=1,2).
$$

В первом приближении ищем решение уравнения Вигнера в виде
$$
f=f^{(1)}=f_0(P)+f_1,
\eqno{(2.1)}
$$
где $f_1$ -- линейная комбинация векторных потенциалов (1.4).

Уравнение Вигнера (1.2) в линейном по величине векторных потенциалов
приближении имеет вид
$$
\dfrac{\partial f}{\partial t}+
v_T{\bf P}\dfrac{\partial f}{\partial {\bf r}}+
\dfrac{iev_T}{c\hbar}{\bf P}\sum\limits_{j=1}^{2}{\bf A}_j
\Big[f_0({\bf P}+\dfrac{{\bf q}_j}{2})-
f_0({\bf P}-\dfrac{{\bf q}_j}{2})\Big]=0.
\eqno{(2.2)}
$$ \bigskip

Здесь
$$
f_0\Big({\bf P}\pm \dfrac{{\bf q}_j}{2}\Big)=\left\{1+
\exp\Big[\Big({\bf P}\pm \dfrac{{\bf
q}_j}{2}\Big)^2-\alpha\Big]\right\}^{-1}.
$$

При подстановке (2.1) и  (1.4) в уравнение (2.2), последнее
распадается на два уравнения

$$
i(v_T{\bf k}_j{\bf P}-\omega_j)\varphi_j{\bf A}_j=-
\dfrac{ievT}{c\hbar}\Big[f_0({\bf P}+\dfrac{{\bf q}_j}{2})-
f_0({\bf P}-\dfrac{{\bf q}_j}{2})\Big],
$$
из которых получаем:
$$
{\bf A}_j\varphi_j=-\dfrac{e{\bf PA}_j}{c\hbar k_T}
\dfrac{f_0({\bf P}+\dfrac{{\bf q}_j}{2})-
f_0({\bf P}-\dfrac{{\bf q}_j}{2})}{{\bf q}_j{\bf
P}_j-\Omega_j},\quad j=1,2.
$$

Итак, функция Вигнера в первом приближении построена:
$$
f_1=-\dfrac{e}{c\hbar k_T}\sum\limits_{j=1}^{2}({\bf PA}_j)
\dfrac{f_0({\bf P}+\dfrac{{\bf q}_j}{2})-
f_0({\bf P}-\dfrac{{\bf q}_j}{2})}{{\bf q}_j{\bf
P}_j-\Omega_j}.
\eqno{(2.3)}
$$

Здесь введены безразмерные параметры
$$
\Omega_j=\dfrac{\omega_j}{k_Tv_T},\qquad \qquad
q_j=\dfrac{k_j}{k_T},
$$
$q_j$ -- безразмерное волновое число,
$k_T =\dfrac {mv_T} {\hbar} $ -- тепловое волновое число, $ \Omega_j$
-- безразмерная частота колебаний векторного потенциала
электромагнитного поля ${\bf A}_j$.

\section{Solution of Wigner equation in second approximation}

Во втором приближении ищем решение уравнения Вигнера (1.2) в
виде (1.3), в котором $f_2$ определяется равенством (1.5).
Получаем следующее уравнение
$$
2i(v_T{\bf k}_1{\bf P}-\omega_1){\bf A}_1^2\psi_1+2i
(v_T{\bf k}_2{\bf P}-\omega_2){\bf A}_2^2\psi_2+
$$
$$
+2i\Big(v_T\dfrac{{\bf k}_1+{\bf k}_2}{2}{\bf P}-
\dfrac{\omega_1+\omega_2}{2}\Big){\bf A}_1{\bf A}_2\psi_0=
$$
$$
=-\dfrac{iev_T}{c\hbar}\sum\limits_{j=1}^{2}({\bf PA}_j)\Big[
f_1\Big({\bf P}+\dfrac{{\bf q}_j}{2}\Big)-f_1\Big({\bf P}-
\dfrac{{\bf q}_j}{2}\Big)\Big]+
$$
$$
+\dfrac{ie^2}{2mc^2\hbar}\sum\limits_{j=1}^{2}{\bf A}_j^2\Big[
f_0({\bf P}+{\bf q}_j)-f_0({\bf P}-{\bf q}_j\big)\Big]+
$$
$$
+\dfrac{ie^2}{2mc^2\hbar}{\bf A}_1{\bf A}_2\Big[f_0\Big({\bf P}+
\dfrac{{\bf q}_1+{\bf q}_2}{2}\Big)-f_0\Big({\bf P}-
\dfrac{{\bf q}_1+{\bf q}_2}{2}\Big)\Big]
\eqno{(3.1)}
$$

В уравнении (3.1)
$$
f_1\Big({\bf P}+\dfrac{{\bf q}_m}{2}\Big)=-\dfrac{e}{c\hbar k_T}
\sum\limits_{j=1}^{2}\Big({\bf P}+\dfrac{{\bf q}_m}{2}\Big){\bf
A}_j\times $$$$ \times\dfrac{f_0\Big({\bf P}+\dfrac{{\bf q}_j}{2}+
\dfrac{{\bf q}_m}{2}\Big)-f_0\Big({\bf P}-\dfrac{{\bf q}_j}{2}+
\dfrac{{\bf q}_m}{2}\Big)}{{\bf q}_j\Big({\bf P}+\dfrac{{\bf q}_m}{2}\Big)-
\Omega_j},\quad m=1,2,
$$
$$
f_1\Big({\bf P}-\dfrac{{\bf q}_m}{2}\Big)=-\dfrac{e}{c\hbar k_T}
\sum\limits_{j=1}^{2}\Big({\bf P}-\dfrac{{\bf q}_m}{2}\Big){\bf
A}_j\times $$$$ \times\dfrac{f_0\Big({\bf P}+\dfrac{{\bf
q}_j}{2}-
\dfrac{{\bf q}_m}{2}\Big)-f_0\Big({\bf P}-\dfrac{{\bf q}_j}{2}-
\dfrac{{\bf q}_m}{2}\Big)}{{\bf q}_j\Big({\bf P}-\dfrac{{\bf q}_m}{2}\Big)-
\Omega_j},\quad m=1,2.
$$

Учтем ортогональность волновых векторов и векторных потенциалов
электромагнитных полей:
$$
{\bf q}_m{\bf A}_j=0,\qquad m,j=1,2.
$$

С помощью этих равенств предыдущие два равенства упрощаются:
$$
f_1\Big({\bf P}+\dfrac{{\bf q}_m}{2}\Big)=-\dfrac{e}{c\hbar k_T}
\sum\limits_{j=1}^{2}{\bf P}{\bf A}_j\times
$$
$$
\times\dfrac{f_0\Big({\bf P}+\dfrac{{\bf q}_j+{\bf q}_m}{2}\Big)-
f_0\Big({\bf P}-\dfrac{{\bf q}_j
-{\bf q}_m}{2}\Big)}{{\bf q}_j\Big({\bf P}+\dfrac{{\bf q}_m}{2}\Big)-
\Omega_j},\quad m=1,2,
$$
и
$$
f_1\Big({\bf P}-\dfrac{{\bf q}_m}{2}\Big)=-\dfrac{e}{c\hbar k_T}
\sum\limits_{j=1}^{2}{\bf P}{\bf A}_j\times
$$
$$
\times\dfrac{f_0\Big({\bf P}+\dfrac{{\bf q}_j-{\bf q}_m}{2}\Big)-
f_0\Big({\bf P}-\dfrac{{\bf q}_j+{\bf q}_m}{2}\Big)}
{{\bf q}_j\Big({\bf P}-\dfrac{{\bf q}_m}{2}\Big)-
\Omega_j},\quad m=1,2.
$$

Теперь составим две разности:
$$
f_1\Big({\bf P}+\dfrac{{\bf q}_m}{2}\Big)-
f_1\Big({\bf P}-\dfrac{{\bf q}_m}{2}\Big)=-\dfrac{e}{c\hbar k_T}
\sum\limits_{j=1}^{2}{\bf P}{\bf A}_j\times
$$
$$
\times\Bigg[\dfrac{f_0\Big({\bf P}+\dfrac{{\bf q}_j+{\bf q}_m}{2}\Big)-
f_0\Big({\bf P}-\dfrac{{\bf q}_j
-{\bf q}_m}{2}\Big)}{{\bf q}_j\Big({\bf P}+\dfrac{{\bf q}_m}{2}\Big)-
\Omega_j}-
$$
$$
-\dfrac{f_0\Big({\bf P}+\dfrac{{\bf q}_j-{\bf q}_m}{2}\Big)-
f_0\Big({\bf P}-\dfrac{{\bf q}_j+{\bf q}_m}{2}\Big)}
{{\bf q}_j\Big({\bf P}-\dfrac{{\bf q}_m}{2}\Big)-
\Omega_j}\Bigg].
$$

Теперь найдем сумму этих разностей:
$$
-\dfrac{iev_T}{c\hbar}\Bigg[{\bf PA}_1
\Big[f_1\Big({\bf P}+\dfrac{{\bf q}_1}{2}\Big)-
f_1\Big({\bf P}-\dfrac{{\bf q}_1}{2}\Big)\Big]+
{\bf PA}_2\Big[f_1\Big({\bf P}+\dfrac{{\bf q}_2}{2}\Big)-
f_1\Big({\bf P}-\dfrac{{\bf q}_2}{2}\Big)\Big]\Bigg]=
$$
$$
=\dfrac{ie^2}{c^2m\hbar}\Bigg[({\bf PA}_1)^2\Bigg[
\dfrac{f_0({\bf P+q}_1)-f_0(P)}{{\bf
Pq}_1-\Omega_1+\dfrac{q_1^2}{2}}-\dfrac{f_0(P)-f_0({\bf P-q}_1)}{{\bf
Pq}_1-\Omega_1-\dfrac{q_1^2}{2}}\Bigg]+
$$
$$
+({\bf PA}_1)({\bf PA}_2)
\Bigg[\dfrac{f_0({\bf P}+\dfrac{{\bf q}_1+{\bf q}_2}{2})-
f_0({\bf P}+\dfrac{{\bf q}_1-{\bf q}_2}{2})}{{\bf Pq}_2-\Omega_2+
\dfrac{{\bf q}_1{\bf q}_2}{2}}-
$$
$$
-\dfrac{f_0({\bf P}-\dfrac{{\bf q}_1-{\bf q}_2}{2})-
f_0({\bf P}-\dfrac{{\bf q}_1+{\bf q}_2}{2})}{{\bf Pq}_2-
\Omega_2-\dfrac{{\bf q}_1{\bf q}_2}{2}}+
\dfrac{f_0({\bf P}+\dfrac{{\bf q}_1+{\bf q}_2}{2})-
f_0({\bf P}-\dfrac{{\bf q}_1-{\bf q}_2}{2})}{{\bf Pq}_1-\Omega_1+
\dfrac{{\bf q}_1{\bf q}_2}{2}}-$$$$-
\dfrac{f_0({\bf P}+\dfrac{{\bf q}_1-{\bf q}_2}{2})-
f_0({\bf P}-\dfrac{{\bf q}_1+{\bf q}_2}{2})}{{\bf
Pq}_1-\Omega_1-
\dfrac{{\bf q}_1{\bf q}_2}{2}}\Bigg]+
$$
$$
+({\bf PA}_2)^2\Bigg[
\dfrac{f_0({\bf P+q}_2)-f_0(P)}{{\bf
Pq}_2-\Omega_2+\dfrac{q_2^2}{2}}-\dfrac{f_0(P)-f_0({\bf P-q}_2)}{{\bf
Pq}_2-\Omega_2-\dfrac{q_2^2}{2}}\Bigg]\Bigg].
$$

Уравнение (3.1) распадается на три уравнения, из которых находим
$$
{\bf A}_j\psi_j=\dfrac{e^2}{2c^2p_T^2}\Bigg[({\bf PA}_j)^2
\Bigg(\dfrac{f_0({\bf P+q}_j)-f_0(P)}{{\bf Pq}_j-\Omega_j+
\dfrac{q_j^2}{2}}\Bigg)+
$$
$$
+{\bf A}_j^2\dfrac{f_0({\bf P+q}_j)-f_0({\bf P-q}_j)}{2}\Bigg]
\dfrac{1}{{\bf Pq}_j-\Omega_j},\qquad (j=1,2),
\eqno{(3.2)}
$$
и
$$
{\bf A}_1{\bf A}_2\psi_0=\dfrac{e^2({\bf PA}_1)({\bf PA}_2)}{2c^2p_T^2}
\Bigg[\dfrac{f_0({\bf P}+\dfrac{{\bf q}_1+{\bf q}_2}{2})-
f_0({\bf P}+\dfrac{{\bf q}_1-{\bf q}_2}{2})}{{\bf Pq}_2-\Omega_2+
\dfrac{{\bf q}_1{\bf q}_2}{2}}-
$$
$$
-\dfrac{f_0({\bf P}-\dfrac{{\bf q}_1-{\bf q}_2}{2})-
f_0({\bf P}-\dfrac{{\bf q}_1+{\bf q}_2}{2})}{{\bf Pq}_2-
\Omega_2-\dfrac{{\bf q}_1{\bf q}_2}{2}}+
\dfrac{f_0({\bf P}+\dfrac{{\bf q}_1+{\bf q}_2}{2})-
f_0({\bf P}-\dfrac{{\bf q}_1-{\bf q}_2}{2})}{{\bf Pq}_1-\Omega_1+
\dfrac{{\bf q}_1{\bf q}_2}{2}}-$$$$-
\dfrac{f_0({\bf P}+\dfrac{{\bf q}_1-{\bf q}_2}{2})-
f_0({\bf P}-\dfrac{{\bf q}_1+{\bf q}_2}{2})}{{\bf
Pq}_1-\Omega_1-
\dfrac{{\bf q}_1{\bf q}_2}{2}}\Bigg]\dfrac{1}{{\bf Pq}-\Omega}+
$$
$$
+\dfrac{e^2{\bf A}_1{\bf A}_2}{2c^2p_T^2}\cdot
\dfrac{f_0({\bf P+q})-f_0({\bf P-q})}{{\bf Pq}-\Omega}.
\eqno{(3.3)}
$$

Здесь
$$
{\bf q}=\dfrac{{\bf q}_1+{\bf q}_2}{2}, \qquad
\Omega=\dfrac{\Omega_1+\Omega_2}{2}.
$$

Таким образом, решение уравнения Вигнера построено и во втором
приближении. Оно определяется равенствами (1.3)--(1.5), в
которых функции $\psi_j (j=1,2)$ и $\psi_0$ определяются
равенствами (3.2) и (3.3).

\section{Electric current in quantum plasma}

Плотность электрического тока согласно его определению равна:
$$
{\bf j}=e\int f {\bf v}\dfrac{2d^3p}{(2\pi \hbar)^3}.
%\eqno{(4.1)}
$$

В работе \cite{Lat1} показано, что в нулевом приближении
электрический ток в квантовой плазме равен нулю:
$$
{\bf j}^{(0)}=e\int f_0(P) {\bf v}\dfrac{2d^3p}{(2\pi \hbar)^3}=0.
$$

Поэтому плотность электрического тока равна
$$
{\bf j}=\dfrac{2ep_T^3v_T}{(2\pi\hbar)^3}\int (f_1+f_2)
\Big({\bf P}-\dfrac{e({\bf A}_1+{\bf A}_2)}{mcv_T}\Big)d^3P.
\eqno{(4.1)}
$$

Равенство (4.1) можно представить в виде:
$$
{\bf j}={\bf j}^{\rm linear}+{\bf j}^{\rm quadr}.
$$

Здесь
$$
{\bf j}^{\rm linear}=\dfrac{2ep_T^3v_T}{(2\pi\hbar)^3}\int f_1
{\bf P}d^3P,
\eqno{(4.2)}
$$
$$
{\bf j}^{\rm quadr}=\dfrac{2ep_T^3v_T}{(2\pi\hbar)^3}\int \Big[f_2
{\bf P}-\dfrac{e({\bf A}_1+{\bf A}_2)}{cp_T}f_1\Big]d^3P.
\eqno{(4.3)}
$$

Итак, электрический ток в квантовой плазме есть сумма двух
слагаемых, линейного и квадратичного. Линейное слагаемое
есть плотность тока, направленного вдоль векторного потенциала
электромагнитного поля (т.е. вдоль вектора напряженности поля).
Оно состоит из членов, пропорциональных первой степени векторных
потенциалов. Квадратичное слагаемое есть плотность тока,
ортогонального линейной плотности  тока. Оно направлено вдоль
волнового вектора. Квадратичное слагаемое состоит из членов,
пропорциональных квадрату векторных потенциалов тока и их
произведению.

Представим линейную часть плотности тока (4.2) в явном виде:
$$
{\bf j}^{\rm linear}=-\dfrac{2ep_T^3v_T}{(2\pi\hbar)^3c\hbar k_T}
\sum\limits_{j=1}^{2}\int {\bf P(PA)}_j
\dfrac{f_0\Big({\bf P}+\dfrac{{\bf q}_j}{2}\Big)-
f_0\Big({\bf P}-\dfrac{{\bf q}_j}{2}\Big)}{{\bf
Pq}_j-\Omega_j}d^3P.
$$

Это векторное выражение имеет одну ненулевую компоненту:
$$
{\bf j}^{\rm linear}=j_y(0,1,0),
$$
где
$$
j_y=-\dfrac{2ep_T^3}{(2\pi\hbar)^3cm}
\sum\limits_{j=1}^{2}A_j \times
$$
$$
\times\int
\dfrac{f_0\Big(P_x+\dfrac{q_j}{2},P_y,P_z\Big)-
f_0\Big(P_x-\dfrac{q_j}{2},P_y,P_z\Big)}{q_jP_x-\Omega_j}P_y^2d^3P.
\eqno{(4.4)}
$$

Здесь введено обозначение
$$
f_0\Big(P_x\pm\dfrac{q_j}{2},P_y,P_z\Big)=\dfrac{1}{1+
\exp\Big[(P_x\pm\dfrac{q_j}{2})^2+P_y^2+P_z^2-\alpha\Big]}.
$$

Далее будем использовать более краткое обозначение:
$$
f_0\Big(P_x\pm\dfrac{q_j}{2}\Big)\equiv
f_0\Big(P_x\pm\dfrac{q_j}{2},P_y,P_z\Big).
$$

Найдем числовую плотность концентрацию частиц плазмы, отвечающую
распределению Ферми---Дирака
$$
N=\int f_0(P)\dfrac{2d^3p}{(2\pi\hbar)^3}=
\dfrac{8\pi p_T^3}{(2\pi\hbar)^3}\int\limits_{0}^{\infty}
\dfrac{e^{\alpha-P^2}P^2dP}{1+e^{\alpha-P^2}}=
\dfrac{k_T^3}{2\pi^2}l_0(\alpha),
$$
где
$$
l_0(\alpha)=\int\limits_{0}^{\infty}\ln(1+e^{\alpha-\tau^2})d\tau.
$$

Осуществим очевидную замену переменных в (4.2) и введем
плазменную (ленгмюровскую) частоту
$$
\omega_p=\sqrt{\dfrac{4\pi e^2 N}{m}}.
$$
Затем  используем связь между числовой плотностью частиц
плазмы (концентрацией), тепловым волновым числом электронов и их
химическим потенциалом
$$
N=\dfrac{1}{2\pi^2}k_T^3l_0(\alpha).
$$

В результате получаем, что выражение для тока (4.4) равно
$$
j_y=-\dfrac{\omega_p^2}{8\pi^2c}\sum\limits_{j=1}^{2}A_j
\times $$$$ \times
\int
\Big(\dfrac{1}{q_jP_x-\Omega_j-q_j^2/2}-
\dfrac{1}{q_jP_x-\Omega_j+q_j^2/2}\Big)f_0(P)P_y^2d^3P,
$$
или
$$
j_y=i\dfrac{\omega_p^2}{8\pi^2}\sum\limits_{j=1}^{2}
\dfrac{E_jq_j^2}{\omega_j}\int
\dfrac{f_0(P)P_y^2d^3P}{(q_jP_x-\Omega_j)^2-q_j^4/4}.
$$

Это выражение плотности поперечного тока сводится к двойному
интегралу:
$$
j_y=\dfrac{i\Omega_p^2 k_Tv_T}{8\pi}\sum\limits_{j=1}^{2}
\dfrac{E_jq_j^2}{\Omega_j}\int\limits_{0}^{\infty}
\dfrac{P^4dP}{1+e^{P^2-\alpha}}\int\limits_{-1}^{1}
\dfrac{(1-\mu^2)d\mu}{(q_jP\mu-\Omega_j)^2-q_j^4/4}.
$$

Здесь
$$
\Omega_p=\dfrac{\omega_p}{k_Tv_T}=\dfrac{\hbar\omega_p}{mv_T^2}
$$
-- безразмерная плазменная частота.

\section{Longitudinal current in quantum plasma}

Заметим, что интеграл от второго слагаемого в (4.3) равен нулю.
Поэтому продольный ток в квантовой, генерируемый двумя электромагнитными
полями, равен:
$$
{\bf j}^{\rm quadr} \equiv {\bf j}^{\rm long}=
\dfrac{2ep_T^3v_T}{(2\pi\hbar)^3}\int f_2{\bf P}d^3P.
\eqno{(5.1)}
$$

Таким образом, продольный ток определяется только вторым
приближением $f_2$ функции распределения.

Векторное равенство (5.1) имеет только одну ненулевую
компоненту: ${\bf j}^{\rm long}=j_x(1,0,0)$, где
$$
j_x=\dfrac{2ep_T^3v_T}{(2\pi\hbar)^3}\int f_2{P_x}d^3P.
\eqno{(5.2)}
$$

Продольный ток (5.2) представим в виде суммы трех слагаемых:
$$
j_x=j_1+j_2+j_0.
\eqno{(5.3)}
$$
Здесь
$$
j_j=A_j^2\dfrac{2ep_T^3v_T}{(2\pi\hbar)^3}\int
P_x\psi_jd^3P\qquad (j=1,2),
\eqno{(5.4)}
$$
$$
j_0=A_1A_2\dfrac{2ep_T^3v_T}{(2\pi\hbar)^3}\int P_x\psi_0d^3P.
\eqno{(5.5)}
$$

Из равенств (5.2)--(5.5) вытекает, что продольный ток
представляет собой сумму трех слагаемых --- токов. Первые два
тока $j_1$ и $j_2$ генерируются соответствующми
векторными потенциалами электромагнитных полей. Их величины
пропорциональны квадратам этих векторных потенциалов. Третий ток,
который назовем перекрестным, генерируется взаимодействием
электромагнитных полей и пропорционален произведению величин
векторных потенциалов.

Представим в явном виде формулы для первых двух токов:
$$
j_j=\dfrac{e^3p_Tv_TA_j^2}{(2\pi\hbar)^3c^2}\int \Bigg[\Big(
\dfrac{f_0(P_x+q_j)-f_0(P)}{q_jP_x-\Omega_j+q_j^2/2}-
\dfrac{f_0(P)-f_0(P_x-q_j)}{q_jP_x-\Omega_j-q_j^2/2}\Big)P_y^2+
$$
$$
+\dfrac{f_0(P_x+q_j)-f_0(P_x-q_j)}{2}\Bigg]
\dfrac{P_xd^3P}{q_jP_x-\Omega_j}.
\eqno{(5.6)}
$$

Преобразуем выражение, стоящее перед интегралом в предыдущей
формуле:
$$
C=\dfrac{e^3p_Tv_T}{(2\pi\hbar)^3c^2}A_j^2=
\dfrac{e^3k_T^3v_T}{8\pi^3c^2p_T^2}A_j^2.
$$
Воспользуемся связью между концентрацией (числовой плотностью),
тепловым волновым числом и химпотенциалом:
$$
N=\dfrac{1}{2\pi^2}k_T^3l_0(\alpha), \quad
l_0(\alpha)=\int\limits_{0}^{\infty}\ln(1+e^{\alpha-\tau^2})d\tau.
$$
Тогда
$$
C=A_j^2\dfrac{2\pi^2e^3Nv_T}{8\pi^3c^2p_T^2l_0(\alpha)}=
A_j^2\dfrac{e\omega_p^2}{16\pi^2c^2p_Tl_0(\alpha)}=
$$
$$
=-E_j^2\dfrac{e\omega_p^2}{16\pi^2l_0(\alpha)p_T\omega_j}=
-E_j^2\dfrac{e\Omega_p^2}{16\pi^2l_0(\alpha)p_T\Omega_j}.
$$
Здесь введены безразмерные частоты
$$
\Omega_p=\dfrac{\omega_p}{k_Tv_T}, \qquad
\Omega_j=\dfrac{\omega_j}{k_Tv_T}\qquad (j=1,2).
$$

Введем  продольно-поперечную проводимость $\sigma_{l,tr}$,
$$
\sigma_{l,tr}=\dfrac{e\hbar}{p_T^2}
\Big(\dfrac{\hbar \omega_p}{mv_T^2}\Big)^2=
\dfrac{e}{k_Tp_T}\Big(\dfrac{\omega_p}{k_Tv_T}\Big)^2=
\dfrac{e\Omega_p^2}{p_Tk_T}.
$$
Тогда
$$
C=-\dfrac{E_j^2
\sigma_{l,tr}k_T}{16\pi^2l_0(\alpha)\Omega_1\Omega_2}=
-\dfrac{E_j^2
\sigma_{l,tr}k_j}{16\pi^2l_0(\alpha)\Omega_j^2q_j}, \quad
j=1,2.
$$

Теперь формула (5.6) может быть представлена как
$$
j_j=J_j\sigma_{l,tr}k_jE_j^2.
\eqno{(5.7)}
$$

В (5.7) $J_j$ -- безразмерная плотность тока,
$$
J_j=-\dfrac{1}{16 \pi^2 l_0(\alpha)q_j\Omega_j}
\int\Bigg[\Big(
\dfrac{f_0(P_x+q_j)-f_0(P)}{q_jP_x-\Omega_j+q_j^2/2}-
\dfrac{f_0(P)-f_0(P_x-q_j)}{q_jP_x-\Omega_j-q_j^2/2}\Big)P_y^2+
$$
$$
+\dfrac{f_0(P_x+q_j)-f_0(P_x-q_j)}{2}\Bigg]
\dfrac{P_xd^3P}{q_jP_x-\Omega_j}.
\eqno{(5.8)}
$$

Интеграл в (5.8) сведем к одномерному. Для этого нам понадобятся
следующие равенства.
Вычислим внутренние
интегралы в плоскости $(P_y,P_z)$, переходя к полярным координатам:
$$
\int f_0(P_x\pm q_j,P_y,P_z)P_y^2dP_ydP_z=
\int\limits_{0}^{2\pi}\int\limits_{0}^{\infty}
\dfrac{\cos^2\varphi \rho^3 d\varphi d\rho}
{1+e^{(P_x\pm q_j)^2+\rho^2-\alpha}}=
$$
$$
=\pi\int\limits_{0}^{\infty}
\dfrac{\rho^3d\rho}{1+e^{(P_x\pm q_j)^2+\rho^2-\alpha}}=
\pi \int\limits_{0}^{\infty}
\rho\ln(1+e^{-(P_x\pm q_j)^2-\rho^2+\alpha})d\rho,
$$
где
$$
\rho=\sqrt{P_y^2+P_z^2}.
$$

Аналогично:
$$
\int f_0(P)P_y^2dP_ydP_z=\int\limits_{0}^{2\pi}\int\limits_{0}^{\infty}
\dfrac{\cos^2\varphi \rho^3 d\varphi d\rho}{1+e^{P_x^2+\rho^2-
\alpha}}=
$$
$$
=\pi \int\limits_{0}^{\infty}
\rho\ln(1+e^{-P_x^2-\rho^2+\alpha})d\rho,
$$
$$
\int f_0(P\pm q_j)dP_ydP_z=\int\limits_{0}^{2\pi}\int\limits_{0}^{\infty}
\dfrac{\rho d\varphi d\rho}{1+e^{(P_x\pm q_j)^2+\rho^2-\alpha}}=
$$
$$
=2\pi\int\limits_{0}^{\infty}
\dfrac{e^{\alpha-(P_x\pm q_j)^2-\rho^2}}
{1+e^{\alpha-(P_x\pm q_j)^2-\rho^2}}\rho d\rho=
\pi \ln(1+e^{\alpha-(P_x\pm q_j)^2}).
$$
Отсюда вытекает, что
$$
\int\limits_{-\infty}^{\infty}\int\limits_{-\infty}^{\infty}
f_0(P)dP_ydP_z=\pi\ln(1+e^{\alpha-P_x^2}).
$$

Введем обозначения:
$$
l(P_x\pm q)=\int\limits_{0}^{\infty}\rho\ln(1+
e^{-(P_x\pm q)^2-\rho^2+\alpha})d\rho,
$$
$$
l(P_x)=\int\limits_{0}^{\infty}\rho\ln(1+
e^{-P_x^2-\rho^2+\alpha})d\rho.
$$

Интеграл от второго слагаемого из (5.8) равен:
$$
\dfrac{1}{2}\int
\dfrac{f_0(P_x+q_j)-f_0(P_x-q_j)}{q_jP_x-\Omega_j}P_xd^3P=$$$$=
\dfrac{\Omega_j}{2q_j}\int\dfrac{f_0(P_x+q_j)-f_0(P_x-q_j)}
{q_jP_x-\Omega_j}d^3P=
$$
$$
=\dfrac{\pi \Omega_j}{2q_j}\int\limits_{-\infty}^{\infty}
\ln\dfrac{1+e^{\alpha-(\tau+q_j)^2}}{1+e^{\alpha-(\tau-q_j)^2}}
\dfrac{d\tau}{q_j\tau-\Omega_j}=
$$
$$
=\dfrac{\pi \Omega_j}{2q_j}\int\limits_{-\infty}^{\infty}
\ln(1+e^{\alpha-\tau^2})\Big(\dfrac{1}{q_j\tau-\Omega_j-q_j^2}-
\dfrac{1}{q_j\tau-\Omega_j+q_j^2}\Big)d\tau=
$$
$$
=\pi q_j\Omega_j\int\limits_{-\infty}^{\infty}
\dfrac{\ln(1+e^{\alpha-\tau^2})d\tau}{(q_j\tau-\Omega_j)^2-q_j^4}=
\dfrac{\pi \Omega_j}{q_j}\int\limits_{-\infty}^{\infty}
\dfrac{\ln(1+e^{\alpha-\tau^2})d\tau}{(\tau-\Omega_j/q_j)^2-q_j^2}.
$$

Вычислим интеграл от первого слагаемого. Имеем:
$$
\int\Big(\dfrac{f_0(P_x+q_j)-f_0(P)}{q_jP_x-\Omega_j+q_j^2/2}-
\dfrac{f_0(P)-f_0(P_x-q_j)}{q_jP_x-\Omega_j-q_j^2/2}\Big)
\dfrac{P_y^2P_xd^3P}{q_jP_x-\Omega_j}=
$$
$$
=\pi \int\limits_{-\infty}^{\infty}
\Big(\dfrac{L(P_x+q_j,P_x)}{q_jP_x-\Omega_j+q_j^2/2}+
\dfrac{L(P_x-q_j,P_x)}{q_jP_x-\Omega_j-q_j^2/2}\Big)
\dfrac{P_xdP_x}{q_jP_x-\Omega_j}.
$$
Здесь
$$
L(P_x\pm q_j,P_x)=l(P_x\pm q_j)-l(P_x)=\int\limits_{0}^{\infty}
\rho\ln \dfrac{1+e^{\alpha-(P_x\pm q_j)^2-\rho^2}}
{1+e^{\alpha-P_x^2-\rho^2}}d\rho.
$$

Рассматриваемый интеграл преобразуем  следующим образом:
$$
\int\limits_{-\infty}^{\infty}
\Big(\dfrac{L(\tau+q_j,\tau)}{q_j\tau-\Omega_j+q_j^2/2}+
\dfrac{L(\tau-q_j,\tau)}{q_j\tau-\Omega_j-q_j^2/2}\Big)
\dfrac{\tau d\tau}{q_j\tau-\Omega_j}=
$$
$$
=\int\limits_{-\infty}^{\infty}
\Big(\dfrac{\tau-q_j/2}{q_j\tau-\Omega_j-q_j^2/2}-
\dfrac{\tau+q_j/2,\tau)}{q_j\tau-\Omega_j+q_j^2/2}\Big)
\dfrac{L(\tau+q_j/2,\tau-q_j/2)}{q_j\tau-\Omega_j}d\tau=
$$
$$
=q_j\Omega_j\int\limits_{-\infty}^{\infty}
\dfrac{L(\tau+q_j/2,\tau-q_j/2)d\tau}{(q_j\tau-\Omega_j)
[(q_j\tau-\Omega_j)^2-q_j^4/4]}d\tau.
$$

Окончательно безразмерная плотность тока равна:
$$
J_j=-\dfrac{1}{16\pi l_0(\alpha)\Omega_j}
\int\limits_{-\infty}^{\infty}\Bigg[
\dfrac{L(\tau+q_j/2,\tau-q_j/2)}{(q_j\tau-\Omega_j)
(q_j\tau-\Omega_j)^2-q_j^4/4]}+$$$$+
\dfrac{\ln(1+e^{\alpha-\tau^2})}
{(q_j\tau-\Omega_j)^2-q_j^4}\Bigg]d\tau.
$$

\section{Crossed current}

Выпишем формулу для вычисления перекрестного тока в явном виде:
$$
j_0=\dfrac{e^3p_T^3v_TA_1A_2}{(2\pi\hbar)^3c^2p_T^2}\int
\Bigg[\dfrac{f_0(P_x+q^+)-f_0(P_x-q^-)}{q_1P_x-\Omega_1+q_1q_2/2}-
$$
$$
-\dfrac{f_0(P_x+q^-)-f_0(P_x-q^+)}{q_1P_x-\Omega_1-q_1q_2/2}+
\dfrac{f_0(P_x+q^+)-f_0(P_x+q^-)}{q_2P_x-\Omega_2+q_1q_2/2}-
$$
$$
-\dfrac{f_0(P_x-q^-)-f_0(P_x-q^+)}{q_2P_x-\Omega_2-q_1q_2/2}\Bigg]
\dfrac{P_y^2P_xd^3P}{qP_x-\Omega}+
$$
$$
+\dfrac{e^3p_T^3v_TA_1A_2}{(2\pi\hbar)^3c^2p_T^2}\int
\dfrac{f_0(P_x+q)-f_0(P_x-q)}{qP_x-\Omega}P_xd^3P.
$$

Перепишем это равенство короче:
$$
j_0=\dfrac{e^3p_Tv_TA_1A_2}{(2\pi\hbar)^3c^2}(J_1-J_2+J_3-J_4+J_5).
\eqno{(6.1)}
$$
Здесь
$$
J_1=\int\dfrac{f_0(P_x+q^+)-f_0(P_x-q^-)}{q_1P_x-\Omega_1+q_1q_2/2}
\dfrac{P_y^2P_xd^3P}{qP_x-\Omega},
$$
$$
J_2=\int\dfrac{f_0(P_x+q^-)-f_0(P_x-q^+)}{q_1P_x-\Omega_1-q_1q_2/2}
\dfrac{P_y^2P_xd^3P}{qP_x-\Omega},
$$
$$
J_3=\int\dfrac{f_0(P_x+q^+)-f_0(P_x+q^-)}{q_2P_x-\Omega_2+q_1q_2/2}
\dfrac{P_y^2P_xd^3P}{qP_x-\Omega},
$$
$$
J_4=\int\dfrac{f_0(P_x-q^-)-f_0(P_x-q^+)}{q_2P_x-\Omega_2-q_1q_2/2}
\dfrac{P_y^2P_xd^3P}{qP_x-\Omega},
$$
$$
J_5=\int\dfrac{f_0(P_x+q)-f_0(P_x-q)}{qP_x-\Omega}P_xd^3P.
$$

Здесь
$$
q=q^+=\dfrac{q_1+q_2}{2},\quad q^-=\dfrac{q_1-q_2}{2}, \quad
\Omega=\Omega^+=\dfrac{\Omega_1+\Omega_2}{2}.
$$

Вычислим внутренние
интегралы в плоскости $(P_y,P_z)$, переходя к полярным координатам:
$$
\int f_0(P_x\pm q^\pm,P_y,P_z)P_y^2dP_ydP_z=
\int\limits_{0}^{2\pi}\int\limits_{0}^{\infty}
\dfrac{\cos^2\varphi \rho^3 d\varphi d\rho}
{1+e^{(P_x\pm q^\pm)^2+\rho^2-\alpha}}=
$$
$$
=\pi\int\limits_{0}^{\infty}
\dfrac{\rho^3d\rho}{1+e^{(P_x\pm q^\pm)^2+\rho^2-\alpha}}=
\pi \int\limits_{0}^{\infty}
\rho\ln(1+e^{-(P_x\pm q^\pm)^2-\rho^2+\alpha})d\rho,
$$
где
$$
\rho=\sqrt{P_y^2+P_z^2}.
$$

Аналогично:
$$
\int f_0(P)P_y^2dP_ydP_z=\int\limits_{0}^{2\pi}\int\limits_{0}^{\infty}
\dfrac{\cos^2\varphi \rho^3 d\varphi d\rho}{1+e^{P_x^2+\rho^2-
\alpha}}=
\pi \int\limits_{0}^{\infty}
\rho\ln(1+e^{-P_x^2-\rho^2+\alpha})d\rho,
$$
$$
\int f_0(P\pm q)dP_ydP_z=\int\limits_{0}^{2\pi}\int\limits_{0}^{\infty}
\dfrac{\rho d\varphi d\rho}{1+e^{(P_x\pm q)^2+\rho^2-\alpha}}=
$$
$$
=2\pi\int\limits_{0}^{\infty}
\dfrac{e^{\alpha-(P_x\pm q)^2-\rho^2}}
{1+e^{\alpha-(P_x\pm q)^2-\rho^2}}\rho d\rho=
\pi \ln(1+e^{\alpha-(P_x\pm q)^2}).
$$

Введем обозначения:
$$
l(P_x\pm q)=\int\limits_{0}^{\infty}\rho\ln(1+
e^{-(P_x\pm q)^2-\rho^2+\alpha})d\rho,
$$
$$
l(P_x)=\int\limits_{0}^{\infty}\rho\ln(1+
e^{-P_x^2-\rho^2+\alpha})d\rho.
$$

Пользуясь приведенными выше равенствами, интегралы $J_1,...,J_5$
сведем к одномерным интегралам:
$$
J_1=\pi\int\limits_{-\infty}^{\infty}\dfrac{l(\tau+q^+)-l(\tau-q^-)}
{q_1\tau-\Omega_1+q_1q_2/2}\cdot \dfrac{\tau d\tau}{q\tau-\Omega},
$$
$$
J_2=\pi\int\limits_{-\infty}^{\infty}\dfrac{l(\tau+q^-)-l(\tau-q^+)}
{q_1\tau-\Omega_1-q_1q_2/2}\cdot \dfrac{\tau d\tau}{q\tau-\Omega},
$$
$$
J_3=\pi\int\limits_{-\infty}^{\infty}\dfrac{l(\tau+q^+)-l(\tau+q^-)}
{q_2\tau-\Omega_2+q_1q_2/2}\cdot \dfrac{\tau d\tau}{q\tau-\Omega},
$$
$$
J_4=\pi\int\limits_{-\infty}^{\infty}\dfrac{l(\tau-q^-)-l(\tau-q^+)}
{q_2\tau-\Omega_2-q_1q_2/2}\cdot \dfrac{\tau d\tau}{q\tau-\Omega},
$$
$$
J_5=2\pi q^+\Omega\int\limits_{-\infty}^{\infty}
\dfrac{\ln(1+e^{\alpha-\tau^2})d\tau}{(q^+\tau-\Omega)^2-{q^+}^4}.
$$

В интегралах $J_1,...,J_4$ числители соответственно равны:
$$
l(\tau+q^+)-l(\tau-q^-)=\int\limits_{0}^{\infty}
\rho\ln\dfrac{1+e^{\alpha-(\tau+q^+)^2-\rho^2}}
{1+e^{\alpha-(\tau-q^-)^2-\rho^2}}d\rho,
$$
$$
l(\tau+q^-)-l(\tau-q^+)=\int\limits_{0}^{\infty}
\rho\ln\dfrac{1+e^{\alpha-(\tau+q^-)^2-\rho^2}}
{1+e^{\alpha-(\tau-q^+)^2-\rho^2}}d\rho,
$$
$$
l(\tau+q^+)-l(\tau+q^-)=\int\limits_{0}^{\infty}
\rho\ln\dfrac{1+e^{\alpha-(\tau+q^+)^2-\rho^2}}
{1+e^{\alpha-(\tau+q^-)^2-\rho^2}}d\rho,
$$
$$
l(\tau-q^-)-l(\tau-q^+)=\int\limits_{0}^{\infty}
\rho\ln\dfrac{1+e^{\alpha-(\tau-q^-)^2-\rho^2}}
{1+e^{\alpha-(\tau-q^+)^2-\rho^2}}d\rho.
$$

Введем в формулу (6.1) плазменную частоту. Тогда формула (6.1)
перепишется в виде:
$$
j_0=\dfrac{e\omega_p^2A_1A_2}{16\pi^2 c^2p_Tl_0(\alpha)}
(J_1-J_2+J_3-J_4+J_5).
\eqno{(6.2)}
$$

В формуле (6.2) перейдем от величин векторных потенциалов
электромагнитных полей к напряженностям электрических полей:
$$
j_0=-\dfrac{e\Omega_p^2E_1E_2}{16\pi^2 l_0(\alpha)\Omega_1\Omega_2p_T}
(J_1-J_2+J_3-J_4+J_5).
\eqno{(6.3)}
$$

Введем  продольно-поперечную проводимость $\sigma_{l,tr}$,
$$
\sigma_{l,tr}=\dfrac{e\hbar}{p_T^2}
\Big(\dfrac{\hbar \omega_p}{mv_T^2}\Big)^2=
\dfrac{e}{k_Tp_T}\Big(\dfrac{\omega_p}{k_Tv_T}\Big)^2=
\dfrac{e\Omega_p^2}{p_Tk_T}.
$$

Теперь формула (6.3) может быть преобразована к виду:
$$
j_0=-\dfrac{\sigma_{l,tr}E_1E_2(k_1+k_2)}
{16\pi^2 l_0(\alpha)\Omega_1\Omega_2(q_1+q_2)}
(J_1-J_2+J_3-J_4+J_5).
\eqno{(6.4)}
$$

Формулу (6.4) перепишем короче:
$$
j_0=J_0\sigma_{l,tr}E_1E_2(k_1+k_2).
\eqno{(6.5)}
$$

В формуле (6.5) $J_0$ -- безразмерная часть плотности
перекрестного тока,
$$
J_0=-\dfrac{1}{16\pi^2 l_0(\alpha)\Omega_1\Omega_2(q_1+q_2)}
(J_1-J_2+J_3-J_4+J_5).
$$

Таким образом, продольная часть тока окончательно  равна:
$$
j_x=\sigma_{l,tr}[E_1^2k_1J_1+E_2^2k_2J_2+E_1E_2(k_1+k_2)J_0].
\eqno{(6.6)}
$$

Если ввести поперечные поля
$$
\mathbf{E}_j^{\bf \rm tr}=\mathbf{E}_j-
\dfrac{\mathbf{k}_j({\bf E}_j{\bf k}_j)}{k_j^2}=
\mathbf{E}_j-\dfrac{\mathbf{q}_j({\bf E}_j{\bf q}_j)}{q_j^2},
$$
то равенство (6.6) можно записать в инвариантой форме
$$
{\bf j}^{\rm long}=\sigma_{l,tr}[({\bf E}_1^{tr})^2{\bf k}_1J_1+
({\bf E}_2^{tr})^2{\bf k}_2J_2+{\bf E}_1^{tr}{\bf E}_2^{tr}
({\bf k}_1+{\bf k}_2)J_0].
$$

\section{Small values of wave numbers}

В случае малых значений волновых чисел величины токов,
пропорциональных квадратам напряженностей электрических полей,
фактически вычислены в нашей работе \cite{Lat7}:
$$
j_j=-\dfrac{e}{8\pi
\omega_j}\Big(\dfrac{\omega_p}{\omega_j}\Big)^2k_jE_j^2=
-\dfrac{e\Omega_p^2}{8\pi \omega_j\Omega_j^2}k_jE_j^2=
-\dfrac{\sigma_{l,tr}}{8\pi \Omega_j^3}k_jE_j^2,\quad q_j\to 0.
$$

Теперь рассмотрим величину перекрестного тока при малых
значениях волновых чисел. Имеем:
$$
f_0(P_x+q)=f_0(P)-g(P)2P_xq+\cdots, \qquad (q\to 0).
$$
Здесь
$$
g(P)=\dfrac{e^{P^2-\alpha}}{(1+e^{P^2-\alpha})^2}=
\dfrac{e^{\alpha-P^2}}{(1+e^{\alpha-P^2})^2}.
$$
Заметим, что
$$
f_0(P+q)-f_0(P-q)=-4g(P)P_xq+\cdots,\qquad (q\to 0).
$$
Следовательно, плотность перекрестного тока равна:
$$
j_0=-\sigma_{l,tr}E_1E_2(k_1+k_2)
\dfrac{q}{4\pi^2l_0(\alpha)(q_1+q_2)\Omega_1
\Omega_2\Omega}\int g(P)P_x^2d^3P.
$$

Интеграл равен:
$$
\int g(P)P_x^2d^3P=\dfrac{4\pi}{3}\int\limits_{0}^{\infty}
\dfrac{P^4e^{P^2-\alpha}dP}{(1+e^{P^2-\alpha})^2}=\pi l_0(\alpha).
$$
Таким образом, плотность перекрестного тока равна:
$$
j_0=-\sigma_{l,tr}E_1E_2(k_1+k_2)\dfrac{1}{4\pi \Omega_1
\Omega_2(\Omega_1+\Omega_2)}.
$$

Из полученных выражений вытекает, что при малых значениях
волновых  чисел для плотности продольного тока получаем:
$$
j_x=-\dfrac{\sigma_{l,tr}}{8\pi}\Big[E_1^2\dfrac{k_1}{\Omega_1^3}+
E_2^2\dfrac{k_2}{\Omega_2^3}+2E_1E_2\dfrac{k_1+k_2}
{\Omega_1\Omega_2(\Omega_1+\Omega_2)}\Big].
$$

Перепишем эту формулу в векторном виде
$$
{\bf j}^{\rm long}=-\dfrac{\sigma_{l,tr}}{8\pi}
\Big[({\bf E}_1^{tr})^2{\bf k}_1\dfrac{1}{\Omega_1^3}+
({\bf E}_2^{tr})^2{\bf k}_2\dfrac{1}{\Omega_2^3}+
2{\bf E}_1^{tr}{\bf E}_2^{tr}({\bf k}_1+{\bf k}_2)
\dfrac{1}{\Omega_1\Omega_2(\Omega_1+\Omega_2)}\Big].
$$

Эта формула в точности совпадает с соответствующей формулой из
нашей работы \cite{Lat10}. Это означает, что при малых значениях
волновых чисел величина плотности продольного тока в
классической и квантовой плазме совпадает.

{\sc Замечание 1.} При вычислении интегралов, входящих в
безразмерные части плотности продольного тока, следует
воспользоваться известным правилом Ландау (см., например,
\cite{Lat7, Lat9}).

{\sc Замечание 2.} Как вытекает из результатов настоящей работы
в квантовой плазме перекрестный ток определяется слагаемым
$J_0$, пропорциональным сумме волновых векторов ${\bf k}_1+{\bf
k}_2$. В \cite{Lat10} показано, что в классической плазме
перекрестный ток есть сумма двух слагаемых $J_{12}$ и $J_{21}$,
пропорциональных соответственно векторам ${\bf k}_1$
и ${\bf k}_2$. Этот факт затрудняет сравнение перекрестных токов
в классической и квантовой плазмах.

\section{Conclusions}

В настоящей работе решена следующая задача: в квантовой плазме
с произвольной степенью вырождения электронного газа, распространяются две
электромагнитные волны с коллинеарными волновыми векторами.

Для описания поведения квантовой плазмы используется кинетическое
уравнение с нелинейным интегралом Вигнера. Такое уравнение
Вигнера с одним векторным потенциалом электромагнитного поля
было построена в нашей работе \cite{Lat1}.

Уравнение Вигнера решается методом последовательных приближений.
Для этого выделены малые параметры задачи. Такими малыми
параметрами являются отношения изменения энергии электронов под
действием электрического поля к тепловой энергии электронов на
двух характерных длинах. Используется квадратичное разложение
функции распределения.

Оказалось, что учет нелинейности электромагнитных полей обнаруживает
генерирование электрического тока, ортогонального к направлению
электрического поля (т.е. направлению известного классического поперечного
электрического тока). Найдена величина поперечного и продольного
электрических токов.

Рассмотрен случай малых значений волновых чисел.
Оказалось, что величина продольного тока в классической и
квантовой плазме совпадает.

В дальнейшем авторы намерены рассмотреть задачи о колебаниях
плазмы и о скин-эффекте с использованием квадратичного по
потенциалу разложения функции распределения.

\end{document}